# Electromotive Force: A Guide for the Perplexed

C. J. Papachristou *,    A. N. Magoulas **

*Department of Physical Sciences, Naval Academy of Greece, Piraeus 18539, Greece*
*E-mail:  papachristou@snd.edu.gr*

** *Department of Electrical Engineering, Naval Academy of Greece, Piraeus 18539, Greece*
*E-mail:  aris@snd.edu.gr*

**Abstract.** The concept of electromotive force (emf) may be introduced in various ways in an undergraduate course of theoretical electromagnetism. The multitude of alternate expressions for the emf is often the source of confusion to the student. We summarize the main ideas, adopting a pedagogical logic that proceeds from the general to the specific. The emf of a "circuit" is first defined in the most general terms. The expressions for the emf of some familiar electrodynamical systems are then derived in a rather straightforward manner. A diversity of physical situations is thus unified within a common theoretical framework.

## 1. Introduction

The difficulty in writing this article was not just due to the subject itself: we had to first overcome some almost irreconcilable differences in educational philosophy between an (opinionated) theoretical physicist and an (equally -if not more- opinionated) electrical engineer. At long last, a compromise was reached! This paper is the fruit of this "mutual understanding".

Having taught intermediate-level electrodynamics courses for several years, we have come to realize that, in the minds of many of our students, the concept of *electromotive force* (*emf* ) is something of a mystery. What is an emf, after all? Is it the voltage of an ideal battery in a DC circuit? Is it work per unit charge? Or is it, in a more sophisticated way, the line integral of the electric field along a closed path? And what if a magnetic rather than an electric field is present?

Generally speaking, the problem with the emf lies in the diversity of situations where this concept applies, leading to a multitude of corresponding expressions for the emf. The subject is discussed in detail, of course, in all standard textbooks on electromagnetism, both at the intermediate [1-9] and at the advanced [10-12] level. Here we summarize the main ideas, choosing a pedagogical approach that proceeds from the general to the specific. We begin by defining the concept of emf of a "circuit" in the most general way possible. We then apply this definition to certain electrodynamic systems in order to recover familiar expressions for the emf. The main advantage of this approach is that a number of different physical situations are treated in a unified way within a common theoretical framework.

The general definition of the emf is given in Section 2. In subsequent sections (Sec.3-5) application is made to particular cases, such as motional emf, the emf due to a time-varying magnetic field, and the emf of a DC circuit consisting of an ideal battery and a resistor. In Sec.6, the connection between the emf and Ohm's law is discussed.

## 2.  The general definition of emf

Consider a region of space in which an electromagnetic (e/m) field exists. In the most general sense, any *closed* path $C$ (or *loop*) within this region will be called a *"circuit"* (whether or not the whole or parts of $C$ consist of material objects such as wires, resistors, capacitors, batteries, or any other elements whose presence may contribute to the e/m field).

We *arbitrarily* assign a positive direction of traversing the loop $C$, and we consider an element $\vec{dl}$ of $C$ oriented in the positive direction. Imagine now a test charge $q$ located at the position of $\vec{dl}$, and let $\vec{F}$ be the force on $q$ at time $t$:

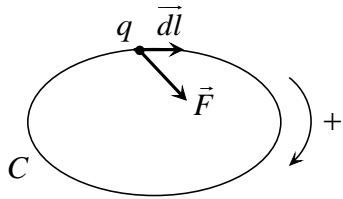

This force is exerted by the e/m field itself, as well as, possibly, by additional energy *sources* (e.g., batteries) that can interact electrically with $q$. The *force per unit charge* at the position of $\vec{dl}$ at time $t$, is

$$\vec{f} = \frac{\vec{F}}{q} \qquad (1)$$

Note that $\vec{f}$ is independent of $q$, since the force by the e/m field and/or the sources on $q$ is proportional to the charge. In particular, reversing the sign of $q$ will have no effect on $\vec{f}$ (although it will change the direction of $\vec{F}$).

We now define the *electromotive force* (*emf*) of the circuit $C$ at time $t$ as the line integral of $\vec{f}$ along $C$, taken in the *positive* sense of $C$:

$$\mathcal{E} = \oint_C \vec{f} \cdot \vec{dl} \qquad (2)$$

Note that the sign of the emf is dependent upon our choice of the positive direction of circulation of $C$: by changing this convention, the sign of $\mathcal{E}$ is reversed.

We remark that, in the *non-relativistic* limit, the emf of a circuit $C$ is the same for all inertial observers since *at this limit* the force $\vec{F}$ is invariant under a change of frame of reference.

In the following sections we apply the defining equation (2) to a number of specific electrodynamic situations that are certainly familiar to the student.

## 3. Motional emf in the presence of a static magnetic field

Consider a circuit consisting of a closed wire $C$. The wire is moving inside a *static* magnetic field $\vec{B}(\vec{r})$. Let $\vec{\upsilon}$ be the velocity of the element $\overrightarrow{dl}$ of $C$ relative to our inertial frame of reference. A charge $q$ (say, a free electron) at the location of $\overrightarrow{dl}$ executes a composite motion, due to the motion of the loop $C$ itself relative to our frame, as well as the motion of $q$ *along* $C$. The total velocity of $q$ relative to us is $\vec{\upsilon}_{tot} = \vec{\upsilon} + \vec{\upsilon}'$, where $\vec{\upsilon}'$ is the velocity of $q$ in a direction parallel to $\overrightarrow{dl}$. The force from the magnetic field on $q$ is

$$\vec{F} = q\,(\vec{\upsilon}_{tot} \times \vec{B}) = q\,(\vec{\upsilon} \times \vec{B}) + q\,(\vec{\upsilon}' \times \vec{B}) \;\; \Rightarrow$$

$$\vec{f} = \frac{\vec{F}}{q} = (\vec{\upsilon} \times \vec{B}) + (\vec{\upsilon}' \times \vec{B})$$

By (2), then, the emf of the circuit $C$ is

$$\mathcal{E} = \oint_{C} \vec{f} \cdot \overrightarrow{dl} = \oint_{C} (\vec{\upsilon} \times \vec{B}) \cdot \overrightarrow{dl} \;+\; \oint_{C} (\vec{\upsilon}' \times \vec{B}) \cdot \overrightarrow{dl}$$

But, since $\vec{\upsilon}'$ is parallel to $\overrightarrow{dl}$, we have that $(\vec{\upsilon}' \times \vec{B}) \cdot \overrightarrow{dl} = 0$. Thus, finally,

$$\mathcal{E} = \oint_{C} (\vec{\upsilon} \times \vec{B}) \cdot \overrightarrow{dl} \tag{3}$$

Note that the wire *need not maintain a fixed shape, size or orientation* during its motion! Note also that the velocity $\vec{\upsilon}$ may vary around the circuit.

By using (3), it can be proven (see Appendix) that

$$\mathcal{E} = -\frac{d\Phi}{dt} \tag{4}$$

where $\Phi = \int \vec{B} \cdot \overrightarrow{da}$ is the magnetic flux through the wire $C$ at time $t$. Note carefully that (4) does not express any novel physical law: it is simply a direct consequence of the definition of the emf!

## 4. Emf due to a time-varying magnetic field

Consider now a closed wire $C$ that is *at rest* inside a *time-varying* magnetic field $\vec{B}(\vec{r}, t)$. As experiments show, as soon as $\vec{B}$ starts changing, a current begins to flow in the wire. This looks impressive, given that the free charges in the (stationary) wire were initially at rest. And, as everybody knows, a magnetic field exerts forces on *moving* charges only! It is also observed experimentally that, if the magnetic field $\vec{B}$ stops varying in time, the current in the wire disappears. The only field that can put an initially stationary charge in motion and keep this charge moving is an *electric* field.

We are thus compelled to conclude that *a time-varying magnetic field is necessarily accompanied by an electric field.* (It is often said that "a changing magnetic field *induces* an electric field". This is somewhat misleading since it gives the impression that the "source" of an electric field could be a magnetic field. Let us keep in mind, however, that the true sources of any e/m field are the electric charges and the electric currents!)

So, let $\vec{E}(\vec{r}, t)$ be the electric field accompanying the time-varying magnetic field $\vec{B}$. Consider again a charge $q$ at the position of the element $\vec{dl}$ of the wire. Given that the wire is now at rest (relative to our inertial frame), the velocity of $q$ will be due to the motion of the charge along the wire only, i.e., in a direction parallel to $\vec{dl}$: $\vec{\upsilon}_{tot} = \vec{\upsilon}'$ (since $\vec{\upsilon} = 0$). The force on $q$ by the e/m field is

$$\vec{F} = q\,[\vec{E} + (\vec{\upsilon}_{tot} \times \vec{B})] = q\,[\vec{E} + (\vec{\upsilon}' \times \vec{B})] \;\Rightarrow$$

$$\vec{f} = \frac{\vec{F}}{q} = \vec{E} + (\vec{\upsilon}' \times \vec{B})$$

The emf of the circuit $C$ is now

$$\mathcal{E} = \oint_C \vec{f} \cdot \vec{dl} = \oint_C \vec{E} \cdot \vec{dl} + \oint_C (\vec{\upsilon}' \times \vec{B}) \cdot \vec{dl}$$

But, as explained earlier, $(\vec{\upsilon}' \times \vec{B}) \cdot \vec{dl} = 0$. Thus, finally,

$$\mathcal{E} = \oint_C \vec{E} \cdot \vec{dl} \qquad (5)$$

Equation (4) is still valid. This time, however, it is not merely a mathematical consequence of the definition of the emf ; rather, it is a true physical law deduced from experiment! Let us examine it in some detail.

In a region of space where a time-varying e/m field $(\vec{E}, \vec{B})$ exists, consider an arbitrary open surface $S$ bounded by the closed curve $C$ :

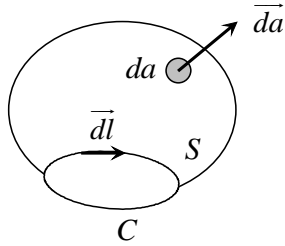

(The *relative* direction of $\vec{dl}$ and the surface element $\vec{da}$, normal to $S$, is determined according to the familiar right-hand rule.) The loop $C$ is assumed stationary relative to the inertial observer; hence the emf along $C$ at time $t$ is given by (5). The magnetic flux through $S$ at this instant is

$$\Phi_m(t) = \int_S \vec{B} \cdot \vec{da}$$

(Note that the signs of $\mathcal{E}$ and $\Phi_m$ depend on the chosen positive direction of $C$.) Since the field $\vec{B}$ is *solenoidal*, the value of $\Phi_m$ for a given $C$ is independent of the choice of the surface $S$. That is, the same magnetic flux will go through *any* open surface bounded by the closed curve $C$.

According to the *Faraday-Henry law*,

$$\mathcal{E} = -\frac{d\Phi_m}{dt} \qquad (6)$$

or explicitly,

$$\oint_C \vec{E} \cdot \vec{dl} = -\frac{d}{dt} \int_S \vec{B} \cdot \vec{da} \qquad (7)$$

(The negative sign on the right-hand sides of (6) and (7) expresses *Lenz's law*.)

Equation (7) can be re-expressed in differential form by using Stokes' theorem,

$$\oint_C \vec{E} \cdot \vec{dl} = \int_S (\vec{\nabla} \times \vec{E}) \cdot \vec{da}$$

and by taking into account that the surface $S$ may be arbitrarily chosen. The result is

$$\vec{\nabla} \times \vec{E} = -\frac{\partial \vec{B}}{\partial t} \qquad (8)$$

We note that if $\partial\vec{B}/\partial t \neq 0$, then necessarily $\vec{E} \neq 0$. Hence, as already mentioned, a time-varying magnetic field is always accompanied by an electric field. If, however, $\vec{B}$ is *static* ($\partial\vec{B}/\partial t = 0$), then $\vec{E}$ is *irrotational*: $\vec{\nabla} \times \vec{E} = 0 \Leftrightarrow \oint \vec{E} \cdot \vec{dl} = 0$, which allows for the possibility that $\vec{E} = 0$.

*Corollary:* The emf around a *fixed* loop $C$ inside a *static* e/m field $\left(\vec{E}(\vec{r}),\ \vec{B}(\vec{r})\right)$ is $\mathcal{E} = 0$ (the student should explain this).

## 5. Emf of a circuit containing a battery and a resistor

Consider a circuit consisting of an ideal battery (i.e., one with no internal resistance) connected to an external resistor. As shown below, the emf of the circuit *in the direction of the current* is equal to the voltage $V$ of the battery. Moreover, the emf in this case represents the *work per unit charge* done by the source (battery).

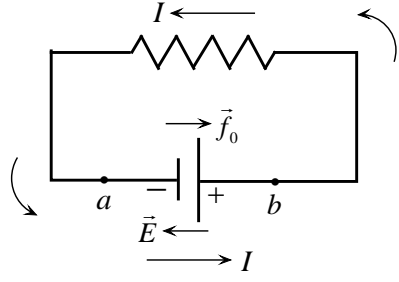

We recall that, in general, the emf of a circuit $C$ at time $t$ is equal to the integral

$$\mathcal{E} = \oint_C \vec{f} \cdot \vec{dl}$$

where $\vec{f} = \vec{F}/q$ is the force per unit charge at the location of the element $\vec{dl}$ of the circuit, at time $t$. In essence, we assume that in every element $\vec{dl}$ we have placed a test charge $q$ (this could be, e.g., a free electron of the conducting part of the circuit). The force $\vec{F}$ on each $q$ is then measured *simultaneously* for all charges at time $t$. Since here we are dealing with a *static* (time-independent) situation, however, we can treat the problem somewhat differently: The measurements of the forces $\vec{F}$ on the charges $q$ need not be made at the same instant, given that nothing changes with time, anyway. So, instead of placing several charges $q$ around the circuit and measuring the forces $\vec{F}$ on each of them at a particular instant, we imagine *a single charge $q$ mak-*ing a complete tour around the loop $C$. We may assume, e.g., that the charge $q$ is one of the (*conventionally positive*) free electrons taking part in the constant current $I$ flowing in the circuit. We then measure the force $\vec{F}$ on $q$ at each point of $C$.

We thus assume that $q$ is a *positive* charge moving *in the direction of the current I*. We also assume that the direction of circulation of $C$ is the *same as the direction of the current* (counterclockwise in the figure). During its motion, $q$ is subject to two forces: (1) the force $\vec{F}_0$ by the source (battery) that carries $q$ from the negative pole $a$ to the positive pole $b$ *through the source*, and (2) the electrostatic force $\vec{F}_e = q\vec{E}$ due to the electrostatic field $\vec{E}$ at each point of the circuit $C$ (both inside and outside the source). The total force on $q$ is

$$\vec{F} = \vec{F}_0 + \vec{F}_e = \vec{F}_0 + q\vec{E} \;\; \Rightarrow \;\; \vec{f} = \frac{\vec{F}}{q} = \frac{\vec{F}_0}{q} + \vec{E} \equiv \vec{f}_0 + \vec{E}$$

Then,

$$\mathcal{E} = \oint_C \vec{f} \cdot \vec{dl} = \oint_C \vec{f}_0 \cdot \vec{dl} + \oint_C \vec{E} \cdot \vec{dl} = \oint_C \vec{f}_0 \cdot \vec{dl} \tag{9}$$

since $\oint_C \vec{E} \cdot \vec{dl} = 0$ for an electrostatic field. However, the action of the source on $q$ is limited to the region between the poles of the battery, that is, the section of the circuit from $a$ to $b$. Hence, $\vec{f}_0 = 0$ *outside* the source, so that (9) reduces to

$$\mathcal{E} = \int_a^b \vec{f}_0 \cdot \vec{dl} \tag{10}$$

Now, since the current $I$ is constant, the charge $q$ moves at constant speed along the circuit. This means that the *total* force on $q$ in the direction of the path $C$ is zero. In the interior of the resistor, the electrostatic force $\vec{F}_e = q\vec{E}$ is counterbalanced by the force on $q$ due to the collisions of the charge with the positive ions of the metal (this latter force does *not* contribute to the emf and is *not* counted in its evaluation!). In the interior of the (ideal) battery, however, where there is no resistance, the electrostatic force $\vec{F}_e$ must be counterbalanced by the *opposing* force $\vec{F}_0$ exerted by the source. Thus, in the section of the circuit between $a$ and $b$,

$$\vec{F} = \vec{F}_0 + \vec{F}_e = 0 \implies \vec{f} = \frac{\vec{F}}{q} = \vec{f}_0 + \vec{E} = 0 \implies \vec{f}_0 = -\vec{E}$$

Equation (10) then takes the final form,

$$\mathcal{E} = -\int_a^b \vec{E} \cdot \vec{dl} = V_b - V_a = V \tag{11}$$

where $V_a$ and $V_b$ are the electrostatic potentials at $a$ and $b$, respectively. This is, of course, what every student knows from elementary e/m courses!

The work done by the source on $q$ upon transferring the charge from $a$ to $b$ is

$$W = \int_a^b \vec{F}_0 \cdot \vec{dl} = q \int_a^b \vec{f}_0 \cdot \vec{dl} = q\,\mathcal{E} \tag{12}$$

[where we have used (10)]. So, the *work of the source per unit charge* is $W/q = \mathcal{E}$. This work is converted into heat in the resistor, so that the source must again supply energy in order to carry the charges once more from $a$ to $b$. This is something like the torture of Sisyphus in Greek mythology!

## 6. Emf and Ohm's law

Consider a closed wire $C$ inside an e/m field. The circuit may contain sources (e.g., a battery) and may also be in motion relative to our inertial frame of reference. Let $q$ be a test charge at the location of the element $\vec{dl}$ of $C$, and let $\vec{F}$ be the total force on $q$ (due to the e/m field and/or the sources) at time $t$. (As mentioned in Sec.2, this force is, classically, a frame-independent quantity.) The force per unit charge at the location of $\vec{dl}$ at time $t$, then, is $\vec{f} = \vec{F}/q$. According to our general definition, the emf of the circuit at time $t$ is

$$\mathcal{E} = \oint_C \vec{f} \cdot \vec{dl} \tag{13}$$

Now, if $\sigma$ is the *conductivity* of the wire, then, by *Ohm's law* in its general form (see, e.g., p. 285 of [1]) we have:

$$\vec{J} = \sigma \vec{f} \qquad (14)$$

where $\vec{J}$ is the *volume current density* at the location of $\vec{dl}$ at time $t$. (Note that the more common expression $\vec{J} = \sigma \vec{E}$, found in most textbooks, is a special case of the above formula. Note also that $\vec{J}$ is measured *relative to the wire*, thus is the same for all inertial observers.) By combining (13) and (14) we get:

$$\mathcal{E} = \frac{1}{\sigma} \oint_C \vec{J} \cdot \vec{dl} \qquad (15)$$

Taking into account that $\vec{J}$ is in the direction of $\vec{dl}$ at each point of $C$, we write:

$$\vec{J} \cdot \vec{dl} = J \, dl = \frac{I}{S} dl$$

where $S$ is the *constant* cross-sectional area of the wire. If we make the additional assumption that, at each instant $t$, the current $I$ is constant around the circuit (although $I$ may vary with time), we finally get:

$$\mathcal{E} = \frac{l}{\sigma S} I = \frac{\rho l}{S} I = I R \qquad (16)$$

where $l$ is the total length of the wire, $\rho = 1/\sigma$ is the *resistivity* of the material, and $R$ is the total resistance of the circuit. Equation (16) is the familiar special form of Ohm's law.

As an example, let us return to the circuit of Sec.5, this time assuming a *non-ideal* battery with internal resistance $r$. Let $R_0$ be the external resistance connected to the battery. The total resistance of the circuit is $R=R_0+r$. As before, we call $V=V_b-V_a$ the potential difference between the terminals of the battery, which is equal to the voltage across the external resistor. Hence, $V=IR_0$, where $I$ is the current in the circuit. The emf of the circuit (in the direction of the current) is

$$\mathcal{E} = I R = I (R_0 + r) = V + I r$$

Note that the potential difference $V$ between the terminals $a$ and $b$ equals the emf only when no current is flowing ($I=0$).

As another example, consider a circuit $C$ containing an ideal battery of voltage $V$ and having total resistance $R$ and total *inductance* $L$:

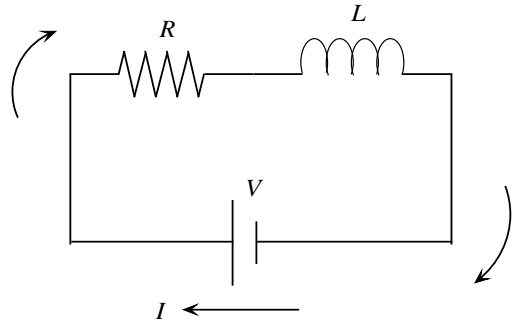

In this case, the emf of *C in the direction of the current flow* is

$$\mathcal{E}(t) = V + V_L = V - L\frac{dI}{dt} = I(t)R$$

To understand why the total emf of the circuit is $V + V_L$, we think as follows: On its tour around the circuit, a test charge $q$ is subject to two forces (ignoring collisions with the positive ions in the interior of the wire): a force *inside* the source, and a force by the *non-conservative* electric field accompanying the time-varying magnetic flux through the circuit. Hence, the total emf will be the sum of the emf due to the (ideal) battery alone and the emf expressed by the Faraday-Henry law (6). The latter emf is precisely $V_L$; it has a nonzero value for as long as the current $I$ is changing.

Some interesting energy considerations are here in order. The total power supplied to the circuit by the battery at time $t$ is

$$P = IV = I^2R + LI\frac{dI}{dt}$$

The term $I^2R$ represents the power *irreversibly lost* as heat in the resistor (energy, per unit time, spent in moving the electrons through the crystal lattice of the conductor and transferred to the ions that make up the lattice). Thus, this power must necessarily be supplied back by the source in order to *maintain* the current against dissipative losses in the resistor. On the other hand, the term $LI(dI/dt)$ represents the energy per unit time required to *build up* the current against the "back emf" $V_L$. This energy is *retrievable* and is given back to the source when the current decreases. It may also be interpreted as energy per unit time required in order to *establish the magnetic field* associated with the current. This energy is "stored" in the magnetic field surrounding the circuit.

## 7. Concluding remarks

In concluding this article, let us highlight a few points of importance:

1. The emf was defined as a line integral of force per unit charge around a loop (or "circuit") in an e/m field. The loop may or may not consist of a real conducting wire, and it may contain sources such as batteries.

2. In the classical (non-relativistic) limit, the emf is independent of the inertial frame of reference with respect to which it is measured.

3. In the case of *purely motional* emf, Faraday's "law" (4) is in essence a mere consequence of the definition of the emf. On the contrary, when a time-dependent magnetic field is present, the similar-looking equation (6) is a true physical law (the Faraday-Henry law).

4. In a DC circuit with a battery, the emf in the direction of the current equals the voltage of the battery and represents work per unit charge done by the source.

5. If the loop describing the circuit represents a conducting wire of finite resistance, Ohm's law can be expressed in terms of the emf by equation (16).

## Appendix

Here is an analytical proof of equation (4) of Sec.3:

Assume that, at time $t$, the wire describes a closed curve $C$ that is the boundary of a plane surface $S$. At time $t'=t+dt$, the wire (which has moved in the meanwhile) describes another curve $C'$ that encloses a surface $S'$. Let $\overrightarrow{dl}$ be an element of $C$ in the direction of circulation of the curve, and let $\vec{\upsilon}$ be the velocity of this element relative to an inertial observer (the velocity of the elements of $C$ may vary along the curve):

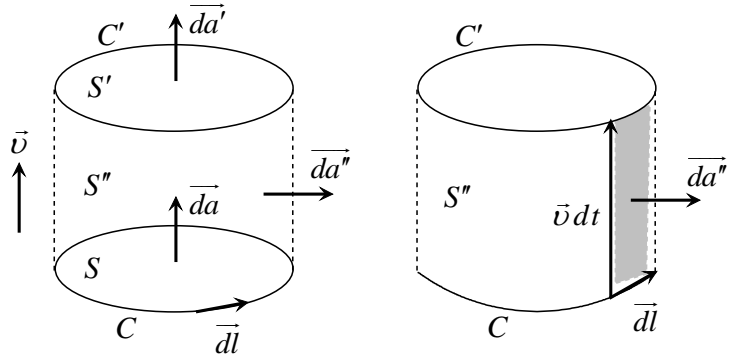

The direction of the surface elements $\overrightarrow{da}$ and $\overrightarrow{da'}$ is consistent with the chosen direction of $\overrightarrow{dl}$, according to the right-hand rule. The element of the side ("cylindrical") surface $S''$ formed by the motion of $C$, is equal to

$$\overrightarrow{da''} = \overrightarrow{dl} \times (\vec{\upsilon}\, dt) = (\overrightarrow{dl} \times \vec{\upsilon})\, dt$$

Since the magnetic field is static, we can view the situation in a somewhat different way: Rather than assuming that the curve $C$ moves within the time interval $dt$ so that its points coincide with the points of the curve $C'$ at time $t'$, we consider two *constant* curves $C$ and $C'$ *at the same instant* $t$. In the case of a *static* field $\vec{B}$, the magnetic flux through $C'$ at time $t'=t+dt$ (according to our original assumption of a moving curve) is the same as the flux through this same curve at time $t$, given that no change of the magnetic field occurs within the time interval $dt$. Now, we note that the open surfaces $S_1=S$ and $S_2= S' \cup S''$ share a common boundary, namely, the curve $C$. Since the magnetic field is *solenoidal*, the same magnetic flux $\Phi_m$ passes through $S_1$ and $S_2$ at time $t$. That is,

$$\int_{S_1} \vec{B} \cdot \overrightarrow{da_1} = \int_{S_2} \vec{B} \cdot \overrightarrow{da_2} \;\; \Rightarrow \;\; \int_{S} \vec{B} \cdot \overrightarrow{da} = \int_{S'} \vec{B} \cdot \overrightarrow{da'} + \int_{S''} \vec{B} \cdot \overrightarrow{da''}$$

But, returning to our initial assumption of a *moving* curve, we note that

$$\int_S \vec{B} \cdot \vec{da} = \Phi_m(t) = \textit{magnetic flux through the wire at time t}$$

and

$$\int_{S'} \vec{B} \cdot \vec{da'} = \Phi_m(t+dt) = \textit{ magnetic flux through the wire at time t+dt}$$

Hence,

$$\Phi_m(t) = \Phi_m(t+dt) + \int_{S''} \vec{B} \cdot \vec{da''} \;\Rightarrow$$

$$d\Phi_m = \Phi_m(t+dt) - \Phi_m(t) = -\int_{S''} \vec{B} \cdot \vec{da''} = -dt \oint_C \vec{B} \cdot (\vec{dl} \times \vec{\upsilon}) \;\Rightarrow$$

$$-\frac{d\Phi_m}{dt} = \oint_C \vec{B} \cdot (\vec{dl} \times \vec{\upsilon}) = \oint_C (\vec{\upsilon} \times \vec{B}) \cdot \vec{dl} = \mathcal{E}$$

in accordance with (3) and (4).

---

[1] http://openeclass.snd.edu.gr/openeclass-2.3.1/modules/document/file.php/TOM6104/EM%20Volume%20PDF.pdf   (in Greek).

[2] http://openeclass.snd.edu.gr/openeclass-2.6.1/modules/document/file.php/TMD107/electromagnetics13.pdf   (in Greek).

[3] One of us (C.J.P.) strongly feels that the 2nd Edition of 1975 (unfortunately out of print) was a much better edition!